# Massive stars as important contributors to two micron light

James E. Rhoads

Princeton University Observatory, Peyton Hall, Princeton, NJ 08544

I: rhoads@astro.princeton.edu

## ABSTRACT

Near infrared light at $2\mu m$ is relatively insensitive to the presence of hot young stars and dust in galaxies, and there has been recent interest in using it as a mass tracer in spiral galaxies. I present evidence that young, cool supergiant stars, whose presence is indicated by strong CO absorption in a $2.36\mu m$ bandpass, dominate the $2\mu m$ light from active star forming regions in the galaxy NGC 1309. The galaxy's quiescent regions, in contrast, do not show evidence of young supergiants. It follows that the $2\mu m$ light comes from different stellar populations in different places, and large changes in the $2\mu m$ surface brightness need not imply correspondingly large features in the galaxy's mass distribution.

*Subject headings:* galaxies: photometry, galaxies: stellar content, infrared: galaxies, galaxies: individual (NGC 1309)

## 1. Introduction

In the quest for a photometric tracer of mass, the near infrared (NIR) bandpasses offer unique advantages. Here both absorption and emission by dust, although present, are small; and the fraction of light from hot young stars is less than at bluer wavelengths. Observations of disk galaxies in increasingly red broadband filters (Schweizer 1976 at $0.64\mu m$; Elmegreen & Elmegreen 1984 at $0.83\mu m$; Rix & Rieke 1993, Rix & Zaritsky 1995 [hereafter RZ] at $2.2\mu m$) have shown that spiral patterns can have large amplitudes in surface brightness at these wavelengths. This has been interpreted as evidence that the spiral pattern involves old stars (e.g., Rix 1993). Because old stars comprise most of the mass in galactic disks, this would mean spirals are large amplitude, nonlinear features.

To test the hypothesis that features in the K band ($2.2\mu m$) light reflect the underlying mass distribution, we can look for spatial variations in the carbon monoxide (CO) index. Rotational-vibrational transitions of CO in stellar atmospheres result in an absorption band at wavelengths $\geq 2.3\mu m$, whose strength can be measured by the photometric index $CO = 2.5 \log_{10}(F_{2.2\mu m}/F_{2.36\mu m}) + c$ (where $F_\lambda$ is the flux in a bandpass with central wavelength $\lambda$). For fixed effective temperature $T_{eff}$ and metallicity $z$, the strength of this band depends strongly on stellar surface gravity $g$. Consequently, it has been used extensively as a luminosity class indicator. The band strength also depends on temperature and metallicity (for tables, see Bell & Briley, 1991), so its interpretation is not entirely straightforward. Nonetheless, if the CO index of two stellar populations differs significantly, the populations themselves must also differ, and it is likely that their mass to $2.2\mu m$ light ratios differ.

I have measured CO indices for both star forming and quiescent regions in the spiral galaxy NGC 1309, and find a large difference that can reasonably be explained if massive, cool supergiant stars dominate the $K$ band light in star forming regions. Thus, $2\mu m$ light is not always a good mass tracer, and spirals with large amplitudes in NIR light do not necessarily have large amplitudes in mass.

### 1.1. The galaxy

NGC 1309 is a nearly face-on Sbc galaxy (de Vaucouleurs et al 1991) that shows prominent bright patches— presumably active regions of star formation— along its northern spiral arm. These patches appear bright blue in Wray's *Color Atlas of Galaxies* (1988) and are also prominent in $K'$ band (RZ). This suggests that young, massive stars dominate the optical light while on the upper main sequence and the $2\mu m$ light during a core helium burning red supergiant phase (see, e.g., figure 8 of Charlot & Bruzual 1991).

NGC 1309's redshift measured from HI observations is $cz = 2135$ km/s (de Vaucouleurs et al 1991). A weak bar may be present in my data; see figure 3. Total or aperture flux measurements have been published at wavelengths of 21 cm (Bottinelli, Gougenheim, & Paturel 1982); 20 cm (Condon et al 1990); 100, 60, 25, and $12\mu m$ (Soifer et al 1989); $K$ and $H$ bands (Devereux 1989); and $V$, $B$, and $U$ bands (Véron-Cetty 1984). RZ obtained $I$ and $K'$ band images, and measure the $K'$ band exponential scale length ($11.1''$) and principal Fourier moments.

## 2. Observations and Analysis

### 2.1. Observations

I observed NGC 1309 in $K'$, 2.22$\mu m$, and 2.36$\mu m$ filters on 18 December 1994 (UT), using the GRIM II camera at $f/5$ (pixel size 0.482″, field of view 123″) on the Apache Point Observatory 3.5 meter telescope. Table 1 gives some parameters of the observations. Exposures of fields 4–5 arcminutes from the galaxy were obtained along with the on-source exposures to allow sky subtraction. Conditions were photometric, and I observed four NIR photometric standard stars (HD 225023, G158-27, HD 40335, & G77-31) from the list of Elias et al (1982).

The data reduction procedures were mostly standard. Individual short exposures were sky subtracted, typically using the two sky frames nearest in time. Flat fields in each band were made by taking the median of all sky frames in that band, subtracting a mean bias frame, and normalizing the result by its median pixel value. The sky subtracted image frames were then divided by these flats.

The only nonstandard step was taken after flat fielding. The NICMOS 3 chip in GRIM II has separate readout electronics for each of its four quadrants. During the NGC 1309 observing session, the median pixel values in each quadrant varied from image to image. The nature of this variation is not clear, nor have I noticed it in subsequent observations with the same equipment. An additive correction was therefore applied to each quadrant, enforcing the constraint that the median pixel values in two annuli without major nonaxisymmetric structures (7.2″ $< r <$ 11″ and 48″ $< r <$ 57.6″) have the same ratio as they would have for a perfect, axisymmetric, face-on exponential disk of scale length 11.1″. I will show later that the systematic error introduced by this correction is smaller than the random errors in our final photometry.

Finally, the corrected image frames were registered and median combined to produce final images. The K′ image is shown in figure 1.

### Table 1

| Filter | central wavelength ($\mu m$) | bandpass FWHM ($\mu m$) | On source integration time | Individual exposure time | Final s/n on nucleus |
|---|---|---|---|---|---|
| $K'$ | 2.124 | 0.337 | 225 s | 15 s | 75 |
| CO cont | 2.22 | 0.089 | 1080 s | 40, 60 s | 75 |
| CO band | 2.36 | 0.094 | 880 s | 40 s | 28 |

## 2.2. Analysis

I next defined masks intended to separate the disk into bright and control regions based on the $K'$ band image. The azimuthally averaged surface brightness profile $\overline{S}_{K'}(r)$ was determined by taking the median pixel value in annuli 8 pixels wide and spaced every 2 pixels, and interpolating for radii between bin centers. This profile was then subtracted from a $K'$ image that had been smoothed with a Gaussian of 2.9 arcsec full width at half maximum (FWHM) to beat down noise. No deprojection was attempted, because NGC 1309's inclination angle is small ($i = 23°$ and $\cos(i) = 0.92$, derived by RZ using the Tully-Fisher relation) and no kinematic determination of the position angle is available. Projection effects would in any case affect only the definitions of masks, and should not introduce any bias in the CO index measurements. Thus my main conclusions are unaffected by projection.

The axisymmetrized radial profiles $\overline{S}(r)$ for the three filters are presented in figure 2, and the nonaxisymmetric part of an unsmoothed $K'$ band image is shown in figure 3.

Two bright region masks were made. Mask 1 was taken as the set $\mathcal{M}_1 := \{(r,\phi) : S_{K'}(r,\phi) > 1.4 \times \overline{S}_{K'}(r)\}$, i.e., any region where the residuals exceeded 40% of the azimuthally averaged surface brightness. Mask 2 was $\mathcal{M}_2 := \{(r,\phi) : S_{K'}(r,\phi) - \overline{S}_{K'}(r) > 27 ADU/pixel \equiv 20.7 K' \,\mathrm{mag/arcsec}^2\}$, i.e., an absolute rather than a relative threshold, with the threshold chosen so that masks 1 and 2 have equal areas. Two control regions (masks 3 and 4) were also made, with $\mathcal{M}_3 := \{(r,\phi) : S_{K'}(r,\phi) < 1.2 \times \overline{S}_{K'}(r)\}$ (residuals < 20% of the axisymmetric light) and $\mathcal{M}_4 := \{(r,\phi) : S_{K'}(r,\phi) - \overline{S}_{K'}(r) < 10.5 ADU/pixel \equiv 21.7 K' \,\mathrm{mag/arcsec}^2\}$ (absolute threshold and the same area as mask 3). In practice, the CO indices in masks 3 and 4 were statistically indistinguishable (differences $< 1\sigma$).

These masks make it possible to combine the high signal to noise (s/n) ratio of conventional aperture photometry with the physical insight of spatially resolved measurements. Define the CO index measured within a region $\mathcal{R}$ to be $CO(\mathcal{R}) = 2.5(\log_{10}(\int_{\mathcal{R}} S_{2.22\mu m}) - \log_{10}(\int_{\mathcal{R}} S_{2.36\mu m})) + c$. Then $CO(\mathcal{M}_1) = 0.195 \pm 0.018$ corresponds roughly to the CO index of the star forming regions, $CO(\mathcal{M}_2) = 0.125 \pm 0.009$ to the spiral arms, and $CO(\mathcal{M}_3) = 0.064 \pm 0.006 \approx CO(\mathcal{M}_4) = 0.059 \pm 0.008$ to the background disk. We can go one step further by defining circular apertures $\mathcal{A}(r_a) = \{(r,\phi) : r < r_a\}$ and for each mask deriving a cumulative profile, $CO(\mathcal{M}_i, r_a) \equiv CO(\mathcal{M}_i \cap \mathcal{A}(r_a))$.

These curves, labeled with their mask number, are shown in figure 5. In the control region, $CO(\mathcal{M}_3, r_a)$ declines from $\sim 0.14$ at the center to $\sim 0.08$ at 12 arcsec, and remains level thereafter. The innermost part of mask 1, a small (7.1 arcsec$^2$) patch at the NW end of the bar, shows a low CO index ($\sim 0$). However, the cumulative CO index for the bright regions converges to a value around 0.19 at the radii where most of the flux in the mask appears. Mask 2 shows CO indices intermediate between the other two masks. The difference between $CO(\mathcal{M}_1)$ and $CO(\mathcal{M}_3)$ is significant at the $6\sigma$ level.

Whatever stellar population gives rise to the axisymmetric light $\overline{S}(r)$ is probably present throughout the disk. The light from mask 1 can therefore be regarded as the sum of $\overline{S}(r)$ and a nonaxisymmetric component. The CO index of the nonaxisymmetric part is a well defined quantity, and for mask 1 is $0.40 \pm 0.055$.

### 2.3. Error budget

The random errors in the CO index measurement are dominated by sky noise in the CO band filter. The error bars in figure 5 are based on combined sky noise for both filters.

Imperfect corrections for the background level fluctuations discussed above may introduce systematic errors. These background fluctuations could be in either the bias or the gain. The galaxy's signal never exceeds 17% of sky. The quadrant fluctuations are on the level of 100 counts, compared with $(1-2) \times 10^4$ sky counts. Thus the fluctuations correspond to either a variation $< \pm 0.01$ in the gain or a variation $\pm 100$ in the bias. I corrected for a bias level problem. If it was in fact the gain, then the fractional error introduced in the measured flux is $< \pm 0.01$. Note further that there are 15-22 individual frames in each final combined image, and since the background fluctuations appear uncorrelated from one exposure to the next the final photometric error should be reduced by an additional factor $\sqrt{N_{exp}} \sim 4$.

Uncorrected color terms in the photometry are a second possible source of systematic error. Two standard stars with known CO indices were observed, and the zero point of the CO index scale determined from the two agrees to within 0.5%. However, color terms cannot be properly modelled, as the two stars have CO indices of $-0.015$ and $-0.008$. Such effects may change all measured CO indices by a factor $\lesssim 1.2$. This may somewhat affect detailed models of the stellar populations in mask 1, but cannot reduce the significance of the difference between $CO(\mathcal{M}_1)$ and $CO(\mathcal{M}_3)$.

### 3. Interpreting the CO Index

The measured CO indices may be used to constrain the properties of NGC 1309's stellar populations. The main astronomical factors that affect the CO index are the star formation rate (SFR), initial mass function (IMF), and metallicity, all integrated appropriately over the history of the galaxy. Furthermore, uncertainties in stellar evolution models (primarily due to convective overshoot and mass loss) cause uncertainties in the CO index calculated for a given star formation history.

We can bypass some of these complications by comparing CO indices for NGC 1309 and for star clusters in the Large Magellanic Cloud (LMC) whose ages have been estimated. Taking 60 clusters with CO photometry from Persson et al (1983) and age estimates from the compilations of Elson & Fall (1985, 1988), we have fitted the linear relation $CO \approx (0.07 \pm 0.02) - (0.06 \pm 0.015) \log(t/\text{Gyr})$, as shown in figure 6. This suggests that light from the underlying disk of NGC 1309 (mask 3) comes from a population with age $\tau \sim 10^9$ yr, while light from the H II regions (mask 1) is dominated by a population with $\tau \sim 10^7$ yr. Since we are now comparing NGC 1309's colors to real star clusters rather than to stellar evolution models, uncertainties in the physics of evolved stars do not enter directly. The smaller uncertainties in main sequence models still do affect our age estimates indirectly, since the LMC cluster age estimates were (mostly) calibrated by main sequence fitting (Elson & Fall 1985, 1988).

The NGC 1309 age estimates are only approximate, for three reasons. First, the scatter in the age-CO relation is large (correlation coefficient $\approx 0.55$). Likely sources of this scatter include cluster-to-cluster variations in metallicity and IMF and measurement errors. For young clusters, statistical fluctuations in the small number of stars currently going through their supergiant phase will also contribute. Second, a linear form is taken for the $CO$-$\log(t)$ relation for pragmatic rather than physical reasons. A full population synthesis model would yield a more complicated curve, probably with several parameters that the data in figure 6 would not constrain strongly. Third, the metallicity and IMF for NGC 1309 may not closely resemble those for the LMC.

Finally, it is possible that hot dust emission contaminates the $2.36\mu m$ CO band photometry. Lester et al (1990) obtained a correction of 0.08 mag (from $CO = 0.22$ to 0.30) for this effect in their observations of M82. Because the NGC 1309 data covers a relatively limited wavelength range, I have not attempted to correct for hot dust emission.

## 4. Discussion

The CO index of the non-axisymmetric component differs from the CO index of the axisymmetric light in NGC 1309. The stellar populations that give rise to the 2 micron light are thus not the same everywhere in the disk, and it seems that the K band surface brightness cannot be used as a reliable estimator of the surface mass density.

The observed CO indices suggest that high mass stars dominate the $K$ band light from star forming regions in NGC 1309's arms, while an older population accounts for much of the light in quiescent interarm regions. The stellar ages $\tau_*$ estimated from comparison with the LMC may be compared with the orbit timescale $\tau_{orb} \sim 10^8$ yr. In star forming regions, $\tau_* \sim 10^7$ yr $< \tau_{orb}$, while in the quiescent disk $\tau_* \sim 10^9$ yr $> \tau_{orb}$. Thus the young stars are still near their formation sites and there is no winding problem.

The NGC 1309 results differ from Rix's (1993) results for M51. Rix found that the $K - CO$ color was almost uniform in M51, being depressed by 0.1 magnitude in only one small patch. A plausible explanation is that the spiral in M51 is driven by M51's tidal interaction with NGC 5195, and therefore involves stars of all ages, while the spiral in NGC 1309 is not strongly driven and primarily reflects star formation activity.

Several future approaches could further elucidate the relation between NIR light and mass distribution in spiral galaxies. First, a wider range of bandpasses could be used to obtain detailed constraints on the stellar populations (cf. Rieke et al 1993) in both arm and interarm regions. Second, spiral galaxies with different environments and structure could be studied, to see if spatial variations in the CO index are less pronounced in grand design spirals with massive companions or bars than in flocculent spirals or galaxies with substantial ongoing star formation. And finally, kinematic information could be used to infer the nonaxisymmetric mass distribution in spirals more directly and to compare it to the light distribution.

Perhaps such work will allow us to calibrate the relation between nonaxisymmetric NIR light and mass. However, until it does, we cannot confidently use broad band NIR light alone as a mass tracer.

I would like to thank Jim Gunn for a seminal remark that set this project in motion; Michael Strauss for discussions and for help getting the data; Ed Fitzpatrick, Bohdan Paczyński, and Stephane Charlot for useful discussions; and David Spergel and Neil Tyson for discussions and for comments on earlier drafts of the manuscript. This research has been supported by NSF grant AST 91-17388 and NASA grant ADP NAG5-269. This research has made use of the Simbad database, operated at CDS, Strasbourg, France.

Figure 1: NGC 1309 in $K'$ band ($\lambda_c = 2.12\mu m$), displayed with a logarithmic intensity transformation. North is down, East is right.

Figure 2: Deviations of the azimuthally averaged radial light profile of NGC 1309 from an exponential disk of scale length 11.1" in each of three filters. Each profile is normalized by its peak intensity. The error bars are $1\sigma$, and the radial bins are independent. The peak intensity in $K'$ band is $16.1\,\mathrm{mag/arcsec^2}$.

Figure 3: Nonaxisymmetric part of the $K'$ band light, displayed with a linear intensity transformation.

Figure 4: Masks used for synthetic aperture photometry of NGC 1309. *a.* Mask 1 (black) corresponds to star forming regions; mask 3 (grey) corresponds to the quiescent disk. *b.* Mask 2 (black) corresponds to the spiral arms; mask 4 (grey) corresponds to the quiescent disk.

Figure 5: Cumulative photometric CO index of NGC 1309 measured in masks 1 (dashed), 2 (dotted), and 3 (solid); results for mask 4 are not shown because they were not significantly different from those for mask 3. The curve for mask $i$ is the CO index of all the flux in the intersection of mask $i$ and a circular aperture of radius $r_a$; thus, different points on the same curve are not independent. The full ranges of CO indices for dwarf, giant, and supergiant stars tabulated by Bell & Briley (1991) is shown at right for comparison. At radii $r \gtrsim 25''$, where the flux within the aperture is large, the CO indices in the different masks converge to significantly different values. This shows that the $2\mu m$ light from these regions comes from different stellar populations. Young supergiants in star forming regions are the most plausible explanation for the difference.

Figure 6: CO index vs. estimated age for a sample of 60 star clusters in the Large Magellanic Cloud, with photometry from Persson et al (1983) and age estimates from Elson & Fall (1985, 1988). The crosshatched regions are the $1\sigma$ error ellipses for the clusters. Also shown are lines for the least squares fit (solid; based on the 2D error ellipses shown) and for $1\sigma$ changes in the fitted line intercept and/or slope (dotted).

Figure 1

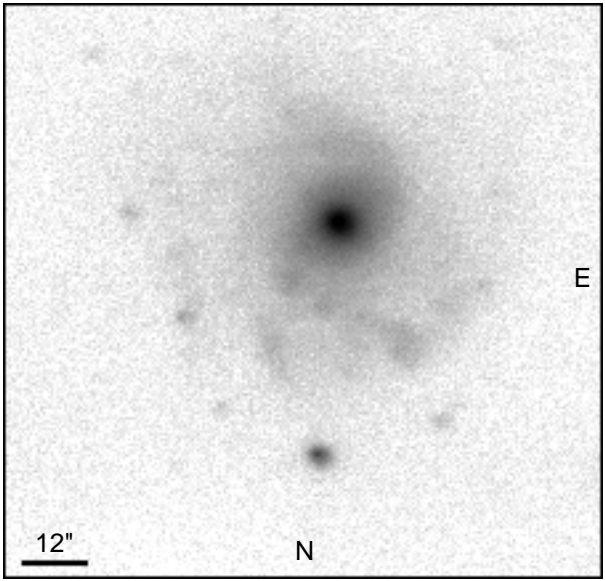

Figure 3

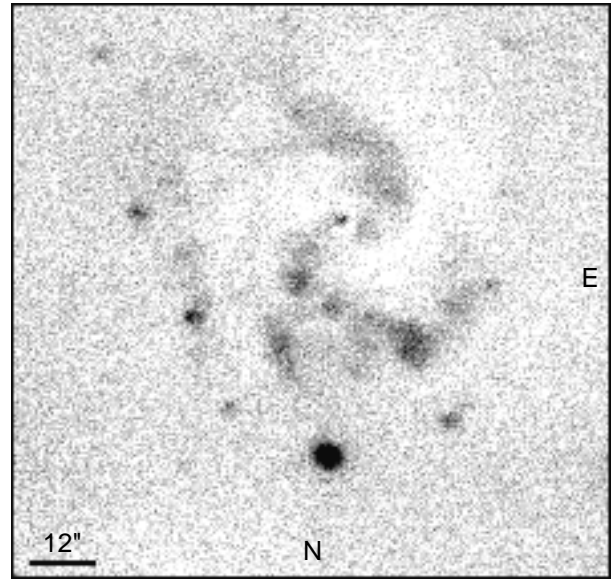

Figure 4a

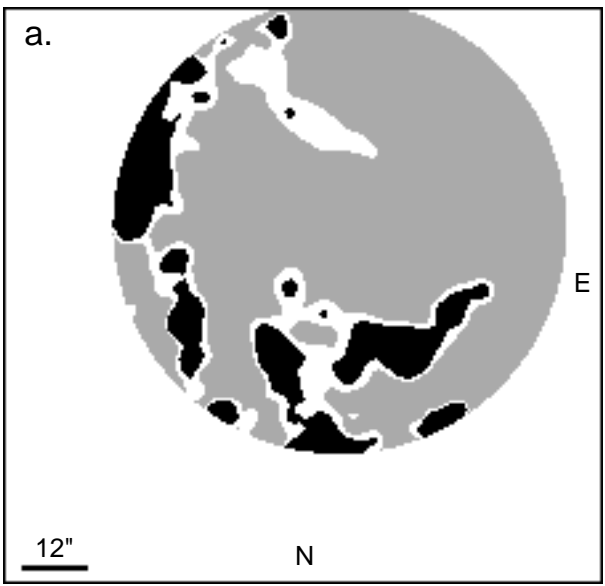

Figure 4b

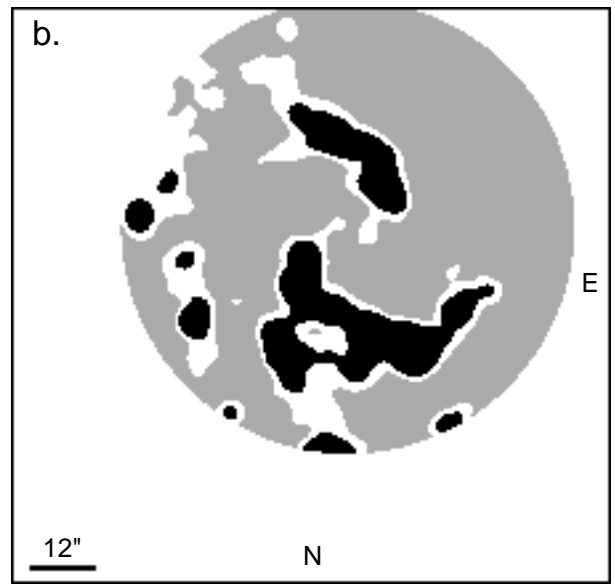

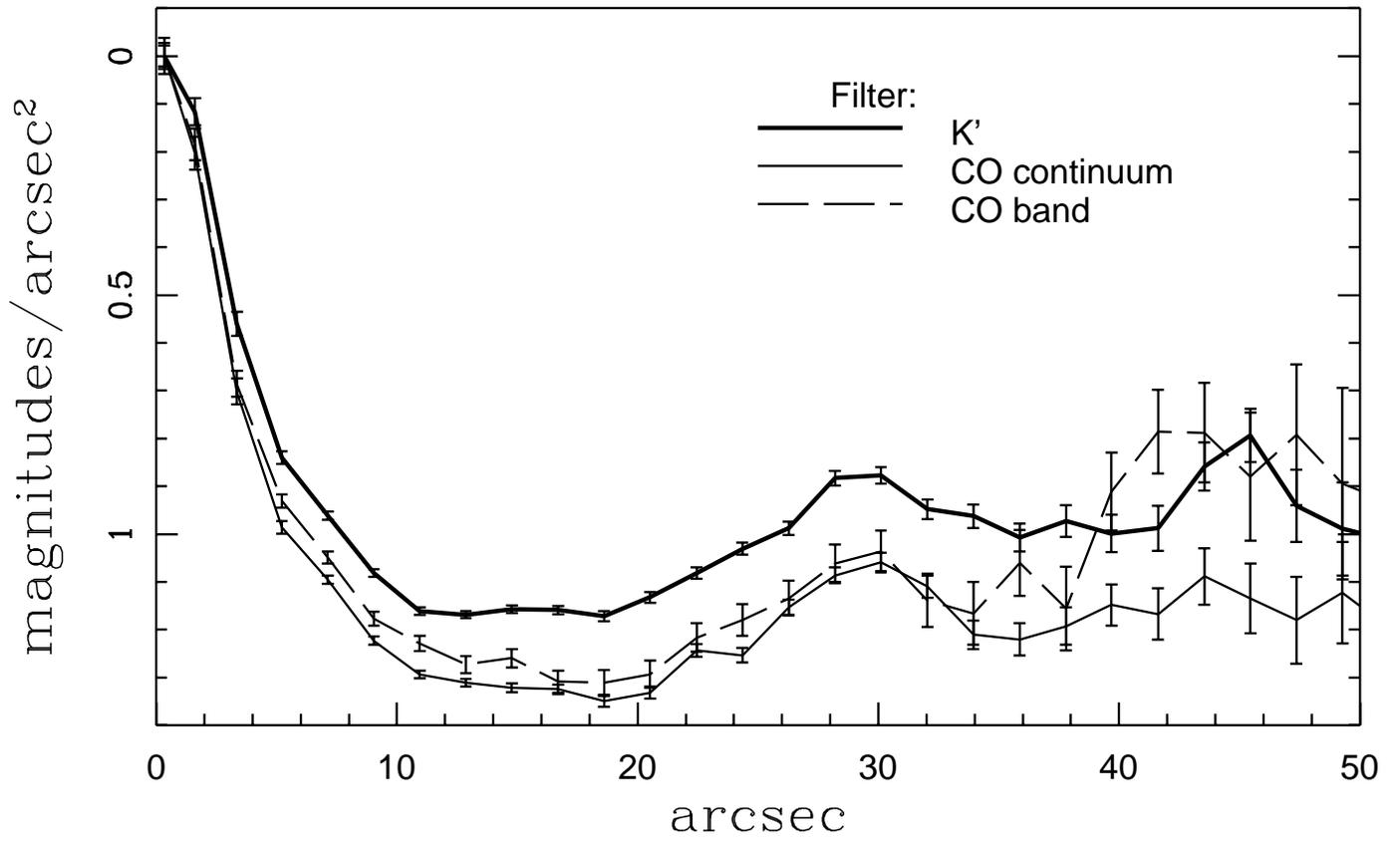

Figure 2

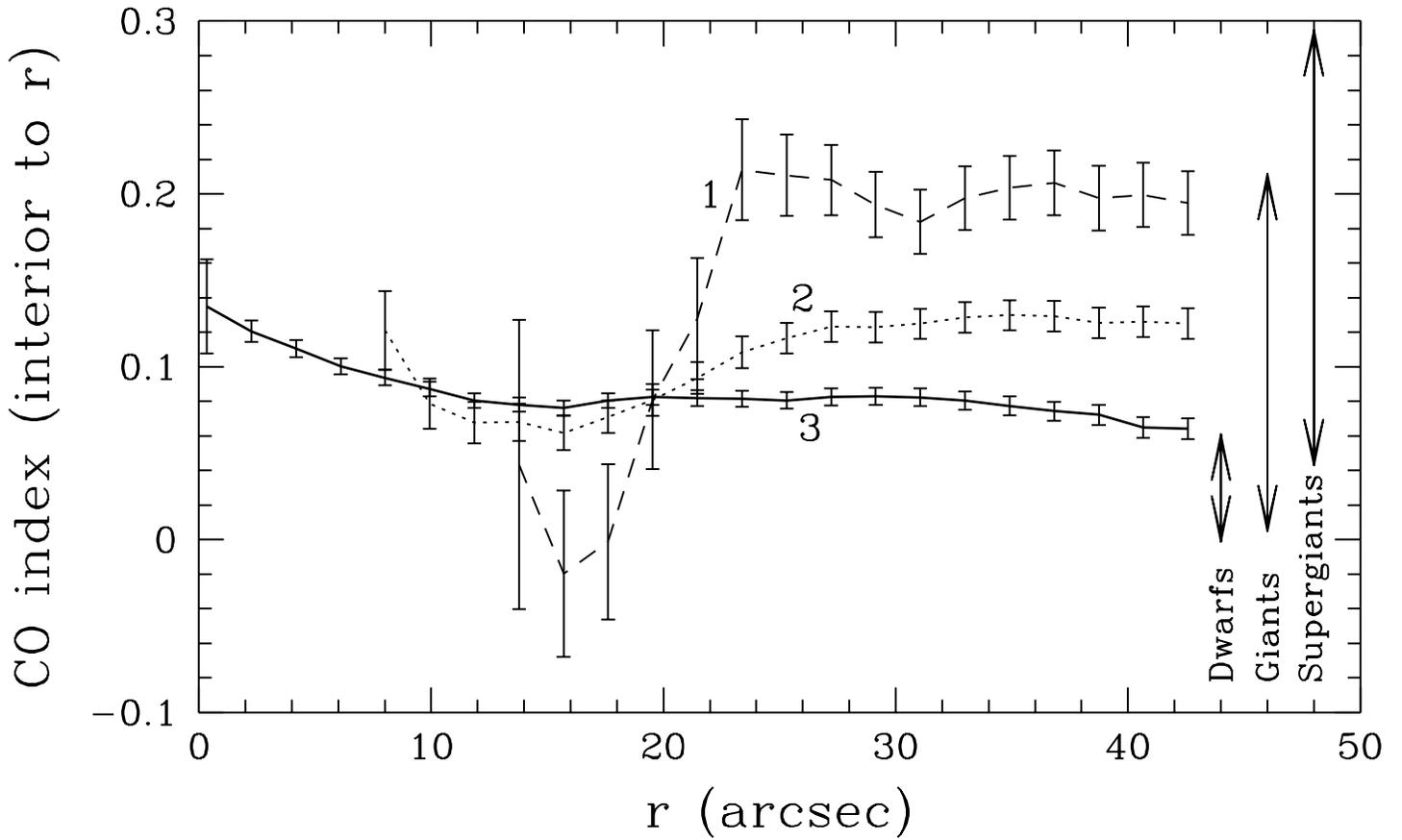

Figure 5

Figure 6

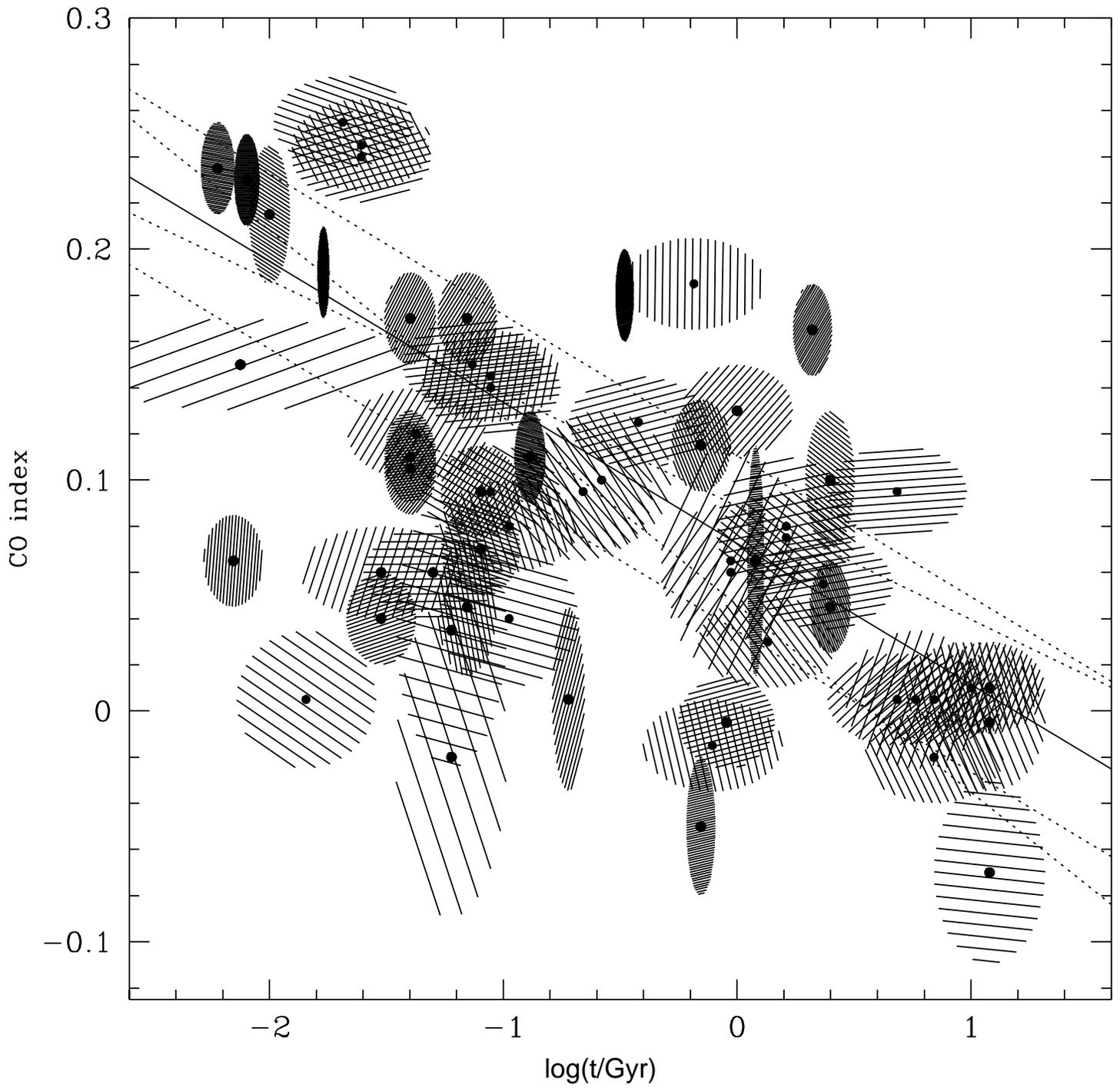